\UseRawInputEncoding
\documentclass[prl,aps,onecolumn,showpacs]{revtex4}

\usepackage{psfrag}
\usepackage{graphicx}
\usepackage{graphics}
\usepackage{bm}
\usepackage{color}
\usepackage{centernot}

\newcommand{\be}{\begin{equation}}
\newcommand{\ee}{\end{equation}}
\newcommand{\ba}{\begin{eqnarray}}
\newcommand{\ea}{\end{eqnarray}}

\newcommand{\dcom}[1]{}
\newcommand{\dnote}[1]{}

\newcommand{\gsim}{\raise.3ex\hbox{$>$\kern-.75em\lower1ex\hbox{$\sim$}}}
\newcommand{\lsim}{\raise.3ex\hbox{$<$\kern-.75em\lower1ex\hbox{$\sim$}}}

\begin{document}

\renewcommand{\thefootnote}{\fnsymbol{footnote}}


\renewcommand{\thefootnote}{\arabic{footnote}}
\setcounter{footnote}{0} \typeout{--- Main Text Start ---}

\title{Un-modelled Pioneer 10/11 clock acceleration from the Schwarzschild metric and Flyby energy increase}
\author{ James C.~ C. Wong}
\affiliation{Department of Electrical and Electronic Engineering, University
of Hong Kong. H.K.}

\date{\today}
\begin{abstract}
When a space-craft reaches a large radial velocity, we expand the time dependent radial distance to linear order in time. In the Schwarzschild metric this linear term in time leads to a significant unmodelled clock acceleration. For Pioneer 10, this acceleration is found to be about $40\%$ of the observed anomalous acceleration. As measured distances are geocentric distances instead of the heliocentric distances in the model, this unmodelled acceleration has an annual oscillation with amplitude close to observations. For Pioneer 11,  we find that the unmodelled acceleration "On-Set"  is caused by the flyby energy Increase to the space-craft's orbit.
\end{abstract}

\pacs{??}

\maketitle

\section{ Introduction}
The Pioneer 10/11 Anomaly poses a challenge to the understanding of gravity \cite{anderson}-\cite{iorio1}. It is found that the radio tracking data (where a ground  based signal is sent to the space-craft which records the frequency and re-transmits the signal back to Earth) from the Pioneer 10 and 11 space-crafts show a small anomalous blue-shifted frequency drift at an uniform changing rate (Eq.(15) of \cite{anderson2})
\begin{equation}
\dot{\mu}=\frac{\mu_{obs}-\mu_s}{\Delta t} =6 \times 10^{-9}Hz/s,
\label{mu}
\end{equation}
where $\mu_{obs}$ is the observed frequency, $\mu_s$ is model frequency, $\Delta t$ is the time lapsed between measurements. 
\\\\
Another interpretation of this frequency drift is that it indicates the existence of a constant acceleration directed towards the Sun \cite{anderson2}, given as
\begin{equation}
a_p=8.74\pm 1.33\times 10^{-10}m/s^2.
\label{ap}
\end{equation}
Pioneer 10 reports a lower acceleration at $a_p =8.09\pm 0.20 \times 10^{-10}ms^{-2}$. For Pioneer 10 at 20AU,  the solar radiation pressure acceleration starts to come below $5\times 10^{-10}m/s^2$,  this is where data collection and analysis start \cite{anderson2}. Pioneer 11 data collection and analysis start much sooner in the mission. After observing the On-Set of the anomalous acceleration of Pioneer 11, data from the Pioneer 10 mission is studied to provide a more complete picture of this acceleration. 
\\\\
Special features in the acceleration data also provide useful constraints on the origin of anomaly. Firstly,  the unmodelled acceleration for Pioneer 11 has an "On-Set" behaviour near Saturn at 9.38 AU \cite{anderson2}. (This On-set behaviour is a topic of investigation \cite{nieto}-\cite{iorio2}.)  Secondly, there is a nearly annual oscillation with an amplitude of $0.215(\pm 0.022) \times 10^{-10}m/s^2$ (subsection IX. C in \cite{anderson2}) which is believed to be due to the Earth's orbiting around the Sun.  
\\\\
In \cite{turyshev}, the authors can account for upto $80\%$ of the anomalous acceleration by a thermal emission term in the line of sight acceleration of the space-craft, which decays exponentially. They further suggest that there is enough uncertainty $(\pm 20\%)$ in the inference of thermal emission so that the anomalous acceleration is argued to be fully modelled. There are however serious reservations about the size of contributions of the thermal recoil force. In \cite{anderson5}, with a simplified space-craft model, Feldman and Andersion find that a $60\%$ contribution from thermal emission is already at high sigma level. Also the thermal radiation pressure does not seem to address directly the Pioneer 10 On-Set behaviour.  Given the continuous growth in solar system explorations, there remain serious interests to find a non-thermal source which can model up to $40\%$ of the acceleration $a_p$ in Eq.(\ref{ap}) .\\\\
From \cite{anderson2} when a planet experiences a small, anomalous,
radial acceleration $a_p$, its orbital radius $r_p$ is perturbed
by an amount
\begin{equation}
\Delta r_p = −\frac{h^6a_p}{(GM⊙)^4} \rightarrow  -\frac{r a_p}{a_N}, 
\label{dr}
\end{equation}
where $M$ is the soloar mass, $h$ is the orbital angular momentum per unit mass
and $a_N$ is the Newtonian acceleration at $r_p$. 
\\\\
From \cite{anderson2}, an universal acceleration $a_p$ would produce a change of the Newtonian planetary orbital radius, which for Earth is $-21 km$ and for Mars is $-76km$. This is contrary to the Viking observations \cite{shapiro}-\cite{anderson3} that both Earth and Mars Newtonian orbital radius are found to be accurate down to $100m$ and $150m$ respectively.  There is a large body of subsequent work which excludes this possibility, see \cite{iorio1}, \cite{anderson5} and references within.
\\\\
Anderson et. al. \cite{anderson2} proposes another possibility which is a clock acceleration such that
\begin{equation}
\mu_{obs} =\mu_s (1+2a_t \Delta t); \:\:\:a_p =ca_t,
\label{at}
\end{equation}
where $c$ is the speed of light with $a_t\sim O(H_0)$ and $H_0$ is the Hubble's constant at present. 
\\\\
There are different ways to obtain clock acceleration \cite{ranada}-\cite{iorio3}.
In this work, we work with the Schwarzschild metric, 
\begin{equation}
c^2d\tau^2=ds^2 = Zc^2dt^2 -\frac{1}{Z}dr^2 -r^2d\Omega^2.
\label{metric2}
\end{equation}
where $\tau$ is the proper time, $t$ is coordinate time  (the Deep Space Network (DSN) frame), $r$ is the radial distance and $\Omega$ is the solid angle with $d\Omega^2 = d\theta^2 +\sin^2\theta d\phi^2$, where $\theta$ and $\phi$ are the polar angles in the spherical symmetric coordinates. 
\\\\
Rearranging the metric in Eq.(\ref{metric2}), we make the approximation that at large distances the solid angle remains constant and obtain
\be 
d\tau^2 =dt^2 Z \bigg(1-\frac{1}{Z^2} \frac{\dot{r}^2}{c^2} \bigg),
\label{taut}
\ee
where
\be
Z=1-\frac{2GM}{rc^2},
\ee
is the Schwarzschild metric factor.
From Eq.(\ref{taut}) we can make a further approximation
\be
d\tau  =dt \bigg( 1-\frac{GM}{rc^2}-\frac{1}{2}\frac{\dot{r}^2}{c^2}\bigg).
\label{tf}
\ee
(Note: for a cosmological constant $\Lambda \sim 10^{-35}s^{-2}$, the Schwarzschild de-Sitter metric factor is
\be
Z=\bigg(1-\frac{2GM}{rc^2}-\frac{\Lambda}{3c^2}r^2\bigg).
\ee
At the relevant scale of our study where $r\sim10^{12}m$, the $\Lambda$ term contributes a value of $O(10^{-28})$ and can be taken to negligible versus the Schwarzschild term which contributes to a value of $O(10^{-9})$. )\\\\
For a fixed wavelength, the clock rates difference can be understood in terms of an effective speed of light $c(t)$ in coordinate time $t$, 
\begin{equation}
c(t)=c\bigg( 1-\frac{GM}{rc^2}-\frac{1}{2}\frac{\dot{r}^2}{c^2}\bigg). 
\end{equation}
This result is well established in Schwarzschild metric \cite{ranada} \cite{kenyon} and in terms of frequency $\mu=c/\lambda$ one has
\be
\mu_{obs}= \mu_s \bigg( 1-\frac{GM}{rc^2}-\frac{1}{2}\frac{\dot{r}^2}{c^2}\bigg).
\label{mu2}
\ee
In the Pioneer considerations, the One-trip Doppler formula for a source moving away from a source/observer at velocity $\dot{r}_p$ is ($s$ for  source)
\begin{equation}
\mu_{obs} =\mu_{s} \bigg(1-\frac{\dot{r}_p}{c}\bigg).
\label{dop}
\end{equation}
The Doppler formula becomes
\begin{equation}
\mu_{obs} =\mu_s \bigg(1-\frac{GM}{r_pc^2} \bigg) \bigg(1-\frac{\dot{r}_p}{c}\bigg),
\label{doppler1}
\end{equation}
here $\dot{r}_p^2/c^2$ is dropped in the presence of $\dot{r}_p/c$. Eq.(\ref{doppler1}) suggests two approaches to model the Pioneer anomaly. 
\\\\ The first approach is to assume that at the time $(t_0)$ of signal measurement, the $\frac{GM}{r_p(t_0) c^2}$ term in Eq.(\ref{doppler1})  remains constant and one looks for any real unmodelled acceleration $a_p$ of the spacecraft, such that
\be
\dot{r}_p(t) = \dot{r}_{p}(t_0) +a_p(t_0)\Delta t,\:\;\: \: \Delta t =t-t_0.
\label{rp}
\ee
This approach is ruled out from the above discussion. We are left with the second approach which is to assume that the unmodelled physical acceleration $a_p$ is negligible and look for an unmodelled clock acceleration within the metric factor $Z^{1/2}$. In this case, we take
\be
r_p(t) =r_p(t_0)+\dot{r}_p(t_0) \Delta t.
\ee
We obtain
\be
c(t)=cZ^{1/2}= c\bigg(1-\frac{GM}{r_p(t_0) c^2} +a_t \Delta t\bigg); \:\;\:\: a_t=\frac{GM}{r_p(t_0) c^2} \bigg(\frac{\dot{r}_p(t_0)}{r_p(t_0)}\bigg).
\label{ct2}
\ee
For simplicity, we write $r_p(t_0)=r_p$ and $\dot{r}_p(t_0) =\dot{r}_p$. \\\\
Given the cosmological constant $\Lambda$, the equation of motion (in an equatorial plane) reads
\begin{equation}
\frac{1}{2}\dot{r_p}^2+\frac{1}{2}\frac{h^2}{r_p^2}-\frac{GM}{r_p}-\frac{\Lambda r_p^2}{6} = E_0
\label{vel}
\end{equation}
where $E_0$ is the particle's total energy.  For simplicity, we work in the Newtonian approximation without the cosmological constant. To book keep, we specify that $R=r_p$ as the Newtonian distance. The flight radial velocity is not given readily by the Pioneer flight path data as they are given in terms of eccentricity $e$ and true anomaly $\varphi$. To proceed, we take the length scale $l$ (where $GMl=h^2$) that is obtainable from $R$, $e$ and $\varphi$ in the orbital equation. We rewrite the energy equation Eq.(\ref{vel}) without the cosmological constant term as
\be
\dot{R}^2 = \frac{2GM}{R}\bigg(1-\frac{l}{2R}+\frac{R}{2l}(e^2-1)\bigg), 
\label{dotr3}
\ee
where $\dot{R}$ is the Newtonian radial velocity. We can further write
\be
\dot{R}^2 = \frac{2GM}{R}\bigg(1 + X\bigg).
\label{X}
\ee
where
\be
X = \frac{R}{2l}(e^2-1)-\frac{l}{2R}=\frac{\bigg(E_0-\frac{h^2}{2R^2}\bigg)}{ \frac{GM}{R}}.
\label{XX}
\ee
On the R.H.S of Eq(\ref{dotr3}), the 2nd term describes the rotational energy and the 3rd term describes the total energy of the particle orbit. The radial speed $\dot{R}$ (from $X$)  can now be obtained from the flight data's orbital eccentricity $e$ and the scale $l$.
\\
\\
Note: the observed maximum secular velocity of Earth's orbital radius is $\dot{r_E} \sim 0.1 r_EH_0$ \cite{ pitjeva} which is 2 order of magnitude smaller than the velocity due to cosmic expansion $H_0r_p (\sim 10r_EH_0)$. We shall ignore its effect in this work.
\section{clock acceleration}
Recall that Eq.(\ref{doppler1}) the Doppler formula for a source moving away from a source/observer at velocity $\dot{r}_p=\dot{R}$ is (the subscript $s$ denotes that the object is from source)
\begin{equation}
\mu_{obs} =\mu_s Z^{1/2}\bigg(1-\frac{\dot{R}}{c}\bigg) =\mu_s \bigg(1-\frac{GM}{Rc^2}+a_t\Delta t\bigg) \bigg(1-\frac{\dot{R}}{c}\bigg) =\mu_s -\mu_s\frac{GM}{Rc^2} -\mu_s\bigg(1-\frac{GM}{Rc^2}\bigg) \frac{\dot{R}}{c} +\mu a_t\Delta t,
\label{muobs}
\end{equation}
here we drop the $ \frac{\dot{R}}{c} (a_t \Delta t)$ term for $\dot{R}/c\ll1$ . The Doppler shift for a round trip is 
\begin{equation}
\mu_D = \mu_{obs} -\mu_{s} = \mu_{D(0)} + A_D,
\end{equation}
where $D$ for Doppler shift which is frequency {\it difference} and not baseline frequency.
\be
\mu_{D(0)} =  -\mu_{s} \frac{2\dot{R}}{c} \bigg(1-\frac{GM}{Rc^2}\bigg)-\mu_s\frac{2GM}{Rc^2},
\label{DS0}
\ee
is the canonical Doppler shift and
\be
A_D(t) =  2\mu_{s}a_t \Delta t,
\label{anDS}
\ee
is the (A for) {\it anomalous blue}  Doppler shift ($a_t>0$) and we have made it the time dependence specific. \\\\
Taking the time derivative of Eq.(\ref{anDS}), where $\dot{R}/c\sim 10^{-4}$, the "one-way" clock acceleration becomes
\begin{equation}
\frac{1}{\mu_s}\frac{dA_D(t)}{dt}=a_t =\frac{GM}{Rc^2} \bigg(\frac{\dot{R}} {R}\bigg) =\frac{a_p}{c} .
\label{acc1}
\end{equation}
We have a simple expression for an effective two-way acceleration 
\be
a_p =2\bigg (\frac{GM}{R^2}\bigg) \bigg(\frac{\dot{R}}{c}\bigg)=2a_N \bigg(\frac{\dot{R}}{c}\bigg)
\label{acc2}
\ee
The one-way acceleration is simply the Newtonian acceleration times a scale factor, which is the ratio $\dot{R}/c$. For closed planetary orbits, the radial velocity 
 is very small compared to the light speed $c$, this acceleration would be negligible. 
\\\\
To see whether $a_p$ from Eq.(\ref{acc2}) has the observed value, we need to find $\dot{R}$ from Eq.(\ref{X}), the orbital eccentricity $e$, true anomaly $\varphi$, the the length scale $l$ using data from Table III in \cite{anderson2} for Pioneer 10 and the orbital equation
\be
R=\frac{l}{1+e \cos \varphi}.
\label{R}
\end{equation}
We obtain $l=2.074\times 10^{20}m$ at $r_p\approx 40AU$, where the one-way clock acceleration $a_t$ is given by
\begin{equation}
a_t=\frac{a_p}{c}  =0.494 \times 10^{-18} s^{-1} 
\end{equation}
The 2-way acceleration is therefore given as
\be
a_p= c\times 0.988\times 10^{-18}m/s^2 =2.96\times 10^{-10}m/s^2.
\ee
At $40AU$, this falls into the range of Pioneer 10 data. This acceleration $a_p$ is about $37\%$ of the the observed $a_p$ value for Pioneer 10 in \cite{anderson2}.
\\\\
Assuming no change of angular momentum over the whole range of distances given in Pioneer 10 data, we further calculate the 2-way acceleration at $r_p =30 AU\:(e =1.22)$ and at $r_p=45AU\:(e =1.733) $ and obtain
\begin{equation}
a_p(30AU)= 3.75 \times 10^{-10}m/s^2; \:\:\:\:a_p(45AU) =2.43 \times 10^{-10}m/s^2
\end{equation}
We note that the calculated $a_p$ value at $30 AU$ is about $43\%$ of the $a_p$ reported by Pioneer 10. This acceleration is consistent with a nearly constant (background) thermal radiation pressure acceleration of $a_p(thermal)\sim 5.2\times 10^{-10}m/s^2$ \cite{turyshev}. We wish to point out that the observed decay in the anomalous acceleration at large distances can be explained by the inverse distance squared of $a_N$ in Eq.(\ref{acc2}). 
\section {The nearly annual oscillation in $a_p$}
In subsection IX.C of \cite{anderson2}, one observes a nearly annual oscillation in $a_p$ that 
$\Delta a_p = 0.215\pm 0.022 \times 10^{-10}m/s^2$. In \cite{anderson5}, the Pioneer 10 oscillation amplitude is measured in Phase I ($40AU-54AU$) and the oscillation amplitude is found dying out in phase III ($\geq 60AU$). The annual nature of the oscillation suggests that this effect is due to the Earth orbiting around the Sun. Since the $R$ used in the above calculations is the distance of the space-craft from the Sun, while in fact we are taking measurements $R_E$ of the distance of the space-craft from the Earth. Approximately, we have $R=R_E(1\pm \frac{1AU}{R_E})$. The use of $R_E$ leads to an uncertainty on the oscillation amplitude, $\Delta a_p$. Using Eq.(\ref{X}) and Eq. (\ref{acc2})
\be
a_p(R)=\frac{2GM}{R^2c}\sqrt{\frac{2GM(1+X)}{R} } = \sqrt{\frac{(2GM)^3(1+X)}{c^2} }  \bigg(\frac{1}{R_E^{5/2}(1\pm \frac{1AU}{R_E} )^{5/2} }\bigg)=a_p(R_E)\mp\Delta a_p(R_E)
\label{osc}
\ee
For $R_E\gg 1AU$,
\be
\frac{1}{R_E^{5/2}(1\pm \frac{1AU}{R_E} )^{5/2}} \approx \frac{1}{R_E^{5/2}} \bigg(1\mp \frac{2.5AU}{R_E}\bigg),
\ee
we obtain 
\be
\Delta a_p(R_E) = a_p(R_E)\bigg(\frac{2.5AU}{R_E}\bigg).
\label{apR}
\ee
Numerically, we see that\\\\
at $R_E=30AU$, $\Delta a_p(R_E)=0.312 \times 10^{-10}m/s^2$;\\
at $R_E=40AU$, $\Delta a_p(R_E) =0.187 \times 10^{-10} m/s^2$, \\
at $R_E=45AU$,  $\Delta a_p(R_E) =0.137 \times 10^{-10} m/s^2$. \\\\
Since the oscillation amplitudes for Pioneer 10 are measured between $40AU \sim 54AU$, our calculation at $40AU$ is reasonably close to the reported oscillation amplitude. Eq.(\ref{apR}) also explains the oscillation amplitude's decay toward $R=60AU$, which is due to both the decay of $a_p(R_E)\propto R_E^{-5/2}$ and $R_E^{-1}$.
\section{The acceleration On-Set }
At 9.38 AU near Saturn, the Pioneer 11 acceleration has an "On-Set" behaviour \cite{anderson2}. Here we show that this acceleration "On-Set" is caused by the energy transfer to the space-craft during its Saturn flyby. From Eq.(\ref{X}), Eq.(\ref{acc2}),  the anomalous acceleration depends on $\dot{R}$ (and  $\sqrt{1+X(e,l)}$). An acceleration "On-Set" requires an abrupt change in the $\sqrt{1+X}$ value from nearly zero before the On-Set to a large fraction after.
\\\\
From the data of Table III in \cite{anderson2}, we pick a point at $R= 22.3AU =3.35\times 10^{12} m$ where Pioneer 11 is an escape orbit far away from the On-Set encounter. We obtain $l=h^2/GM$ (in units of AU) by the orbital equation
\be
l=R (1+e\cos\varphi) =4.4 \times 10^{12}m=29.3 AU,
\ee
and the total energy per unit mass
\be
E_0=\frac{ (e^2-1)GM }{2l}=53.95\:km^2/s^2.
\ee
Before we can evaluate $X$ (and $\dot{R}$), there is a complication. The large flyby energy increase of the space-craft is initially given in terms of rotational energy in the Sun-Saturn plane.  If we choose to describe the Pioneer 11 orbit in the Sun-Saturn plane, from Fig. 2 of \cite{nieto2},  we see that the flight orbit has a significant "out-of-plane" velocity component $v_z$ along the angular momentum vector of the Sun-Saturn plane. The space-craft's projected path on the Sun-Saturn plane will have a reduced total energy, by subtracting out the kinetic energy due to the velocity component $v_z$ from the system's total energy. We need to estimate  $v_z$ of Pioneer 11 from the Saturn flyby data documented in \cite{nieto2}-\cite{nieto3}, assuming that $R$ recorded in the flight path is close enough to its projection on the Sun-Saturn plane.
\\\\
The incoming Pioneer 11 reaches a near Saturn region at $R=7.509AU$ in August 1978. Using data in \cite{nieto3}, we obtain the corresponding $l=6.34AU$ from eccentricity at $e=0.9919$.
Pioneer 11 eventually reaches Saturn in 1979 September ($r_p\approx 9.38 AU$) with an approximated closest approach of $24000\:km$ coming from a trajectory below the Saturn ring, which indicates the existence of  a non-orbital plane velocity component.\\\\
During the flyby encounter there is an increase of space-craft's total energy and angular momentum (and rotational energy) from the Saturn orbit, which is described by the Jacobi Equation (cf. Eq.3 \cite{nieto2}) as follows, where $M$ is solar mass
\be
\bigg(\frac{1}{2}\dot{R}^2+\frac{h^2}{2R^2} +\frac{1}{2}v_z^2 \bigg) -\frac{GM}{R}-\frac{GM_{sat}}{R_s} =E_0 =h\sqrt{\frac{GM}{R^3}}-\frac{C}{2},
\label{jacobi}
\ee
where $C$ is the Jacobi constant, $M_{Sat}$ is the Saturn mass and $R_s$ is Pioneer 11's distance from Saturn (centre of mass). In the co-rotating plane of Saturn (where the Sun-Saturn line is fixed), the kinetic energy (per unit mass) in the bracket of Eq.(\ref{jacobi}) depends on the radial velocity squared, the rotational energy and the out-of-plane velocity squared. There is a special condition that near the flyby, the radial velocity becomes very small. In this case, the data of energies can yield a good estimate of $v_z$.
\\\\
From \cite{nieto2} Fig. 7, we estimate from reading the charts that the in-coming space-craft has a total energy $E_0= -27(\pm2) km^2/s^2$ with the potential energy due to Saturn at $-20 km^2/s^2$, and the potential energy due to the Sun is $-94.5 km^2/s^2$ (at 9.38AU), so that the total kinetic energy per unit mass is taken as $87.5 km^2/s^2$.\\\\
From Eq.(\ref{jacobi}), at $-20\:hr$ and taking a value of $J=-C/2=-121.5\:km^2/s^2$ from \cite{nieto2} Fig. 7, we can estimate the space-craft's $l$ value at the flyby distance $R=9.38AU$ by rearranging the last equation of Eq.(\ref{jacobi}),
\be
l=R\bigg( \frac{E_0+\frac{C}{2}}{\frac{GM}{R}}\bigg)^2 =R \bigg(\frac{94.5}{94.5}\bigg)^2=R=9.38AU.
\label{lb4flyby}
\ee
(From \cite{nieto3}, where the projected true anomaly is near $\varphi \approx 90^{0}$ which is  consistent with Eq. (\ref{lb4flyby}).) From $l$, the orbital rotational energy of the spacecraft is given by
\be
\frac{h^2}{2R^2}=\frac{1}{2} \frac{GM}{R} =47.2 km^2/s^2.
\ee
From \cite{nieto3}, the space-craft takes a circular path approaching Saturn before the energy increase. For a period before the flyby, the space-craft kinetic energy has only a small radial velocity component so that one can estimate that
\be
\Delta E_z=\frac{1}{2} v_z^2 \leq (87.5-47.2) km^2/s^2 =40.3 km^2/s^2.
\ee
After the Saturn flyby, the space-craft has a kinetic energy (and thus total energy) increase by $80.0\pm 3.0\:km^2/s^2$.  From the Jacobi Equation Eq.(\ref{jacobi}), one sees that the kinetic energy increase is attributed to the increase of angular momentum only. The total angular momentum increase after flyby is represented by
\be
l=R\bigg(\frac{94.5+80}{94.5}\bigg)^2 =31.9 AU
\ee
This is larger than the $l=29.3AU$ at 22.3 AU. (This is noted in \cite{nieto2} that as the total energy and angular momentum are not constant of motion in this three-body consideration and they will reduce asymptotically to two-body trajectory values as the space-craft leaves the influence of the Saturn potential.)
\\\\
On the Sun-Saturn plane, Eq.(\ref{XX}) should be written in reduced total energy as
\be
X= (E_0-\Delta E_z) \bigg(\frac{GM}{R}\bigg)^{-1}-\frac{l}{2R},
\label{x2}
\ee
where reduced total energy used is obtained by subtracting the "out-of-plane" kinetic energy from the total energy. We are now calculate the unmodelled acceleration.
We recall that before the flyby, the total energy $E_0=-27km^2/s^2$ which is negative, $1+X \ll 1$ and the $a_p$ value is suppressed by the factor $\sqrt{1+X}$ (Eq.(\ref{dotr3}) and Eq.(\ref{acc2})). After the flyby, $E_0$ has increased by $80\:km^2/s^2$ and the reduced total energy in the orbital plane becomes positive, one expects to see significant impact on $a_p$ after the flyby. For the purpose of seeing the acceleration On-Set, we follow the flow of the value $\sqrt{1+X}$ in regions before and after the flyby. 
\\\\
At an incoming point with $R=7.5AU$, $l=6.34AU$, taking $\Delta E_0=40.3 km^2/s^2$ and the potential due to Saturn is negligible at this distance, we arrive at
\be
X(7.5AU)= -0.991; \:\;\: \sqrt{1+X(7.5AU)} =0.095,
\ee
which leads to a suppression of $a_p$. The estimated acceleration $a_p (7.5AU)\sim 10\times 10^{-10}m/s^2$ provides the correct order of magnitude. However, had we taken $E_0=-28km^2/s^2$, we have $\sqrt{1+X}=0$ where there is no $a_p$. Given the sensitivity of our energy estimate to the value of $\sqrt{1+X}$ at close to flyby distance, we should only take the calculated value of  $\sqrt{1+X}$ and $a_p(7.5AU)$ to indicate that $\dot{R}$ is criticallly close to zero (on a circular path) near the flyby region.\\\\
Immediately before the flyby at $R=9.38AU$, here we need to include the Saturn potential energy $-20 km^2/s^2$ to that from the sun. The total energy (from energy transfer) needs to start increasing from $-27km/s^2$ to ensure that
\be
 \sqrt{1+X(9.38AU)}\geq 0,
\ee
which maximally suppresses $a_p$. This total energy increase behaviour is observable in \cite{nieto2} Fig. 7.
\\\\
Immediately after the flyby, the increase in the reduced total energy is in rotational energy form w.r.t the rotating plane. In Eq.(\ref{x2}) the two energy increases cancel and there is no net change in $X$, so that change of clock acceleration remains small. It is until after the flyby when the angular momentum comes down to its aymptotic value and some of the total energy increase is channelled into radial kinetic energy increase that we obtain a boost in the radial velocity (and $a_p$).  We go to a time long after the flyby, where the different energies and also the $l$ value become settled. The reduced total energy at this point is 
\be
E_0-\Delta E_z=-27km/s^2+80km^2/s^2-40.3 km^2/s^2=12.7>0.
\ee
From Eq.(\ref{x2}) $X$ will increase as $R$ increases while the $\frac{l}{2R}$ term will decrease as $R$ increases for fixed $l$. 
To see this numerically, at $R=22.3AU$, we obtain
\be
X= -0.337;\:\:\: \sqrt{1+X}=0.813,
\ee
comparing with the values at $7.5AU$, the value $\sqrt{1+X}$ here provides much less suppression for radial velocity and the acceleration is given by
\be
a_p(22.3 AU)=  5.60\times 10^{-10}m/s^2.
\ee
At this distance, we expect that the thermal radiation decay contribution to $a_p$ becomes important.
\\
\\At the distance $R=15AU$, we assume that the space-craft has already taken up the new total energy $E_0$, its angular momentum is already settled at its value at $R=22.3AU$ and it is sufficiently escaped from the Saturn potential, we find that 
\be
X= -0.762;\:\:\sqrt{1+X}=0.488,
\ee
so that at this distance $\sqrt{1+X}$ no longer strongly suppresses its radial velocity and the factor $\sqrt{1+X}$ at above this distance is expected to behave smoothly. At this distance, we obtain a significant acceleration value increase at
\be
a_p (15AU) =9.27\times 10^{-10}m/s^2.
\ee
This $a_p$ remains within the range of the observation in Fig. 7 \cite{anderson2}, but the room for thermal acceleration decay is small. Here the nearly annual oscillation amplitude provides an error bar of $\pm 16\%$. We see that the "On-Set" of the clock acceleration is closely connected to the flyby energy transfer. On the Sun-Saturn plane across the flyby, the Pioneer 11's trajectory around the Sun moves from a nearly circular orbit with small radial velocity to a hyperbolic orbit  with a large radial velocity. This is designed to take the space-craft away from orbiting the sun towards the edge of the solar system.
\section{conclusion}
The time dependence of a space-craft's radial distance is modelled by a taylor expansion around the time when the signal is measured. For Pioneer 10, when the radial velocity becomes large, the linear term in time increment, through clock acceleration, leads to an unmodelled (one-way) acceleration, expressible as the Newtonian acceleration times the ratio $\dot{R}/c$. Using flight orbital parameters, this clock acceleration is found to have values upto $40\%$ of the observations. This result supports the contention of Feldman and Anderson in \cite{anderson5}.
In our calculations, the radius distance data used are geocentric while the modelled calculations require heliocentric radius, which could differ from the geocentric radius by upto $\pm 1AU$ per year. This leads to an acceleration oscillation with an amplitude which matches observations at $40AU$ and holds a decay pattern towards large distances.
For Pioneer 11, the unmodelled acceleration data shows an On-Set event which coincides with a gravity assist flyby near Saturn. We use the Saturn flyby energies data together with the space-craft's orbital parameters to obtain the unmodelled acceleration close to the flyby region, the resulting acceleration is consistent with the "On-Set" observations of Pioneer 11's plot in \cite{anderson2}. The physical reason for the On-Set is clear. The space-craft's orbit projection on the Sun-Saturn rotating plane before the flyby is elliptical, its total energy is negative and the radial velocity is small. After the flyby the space-craft receives a major energy increase which turns its total orbital energy to positive, the space-craft moves into an hyperbolic escape orbit. This total energy increase causes a boost in its radial velocity that amplifies the clock acceleration which would be hidden when the ratio $\dot{R}/c$ is small.  

\end{document}